\documentclass[11pt]{article}
\usepackage{a4wide}
\usepackage{amssymb}
\usepackage{graphicx}
\usepackage{subfigure}
\usepackage{epsfig}
\usepackage[tbtags]{amsmath}
\usepackage{exscale}

\begin{document}

\newcommand{\be}{\begin{equation}}
\newcommand{\ee}{\end{equation}}
\newcommand{\bea}{\begin{eqnarray}}
\newcommand{\eea}{\end{eqnarray}}
\newcommand{\md}{{\mathrm{d}}}

\begin{center}{\bf{Ferminonic edge states and new physics}}

\bigskip

{T.R. Govindarajan$^a$\footnote{e-mail address: {\tt trg@cmi.ac.in and trg@imsc.res.in}} and Rakesh Tibrewala$^b$\footnote{e-mail address: {\tt rtibs@cts.iisc.ernet.in}}}

\bigskip

$^a${\it Chennai Mathematical Institute,}\\
{\it Kelambakkam 603103, India} \\
\vspace{0.5em}
$^b${\it Center for High Energy Physics,}\\
{\it Indian Institute of Science, Bangalore-560012, India}

\end{center}

\begin{abstract}
We investigate the properties of the Dirac operator on manifolds with boundaries in presence of the Atiyah-Patodi-Singer boundary condition. An exact counting of the number of edge states for boundaries with isometry of a sphere is given. We show that the problem with the above boundary condition can be mapped to one where the manifold is extended beyond the boundary and the boundary condition is replaced by a delta function potential of suitable strength. We also briefly highlight how the problem of the self-adjointness of the operators in the presence of moving boundaries can be simplified by suitable transformations which render the boundary fixed and modify the Hamiltonian and the boundary condition to reflect the effect of moving boundary.

\end{abstract}

\section{Introduction}

Wolfgang Pauli once remarked `God made the bulk, but the 
devil invented the surfaces' \cite{PauliQuote}. The truth of the statement cannot be over emphasized since the presence of boundaries and boundary conditions makes the analysis of a problem much more non-trivial.
Manifolds with boundaries are ubiquitous in physics - from 
topological insulators \cite{KaneRevTI} and quantum Hall systems \cite{HalperinQHE, WenFQHE} in 
condensed matter physics to black hole horizons and the 
cosmological horizon in general relativity \cite{AnninosHorizonBoundary, EardleyBHboundary}. 
The dynamics of the system of 
interest is usually encoded in the form of suitable differential 
operators 
on these manifolds. An analysis of the  
differential operators on manifolds with boundaries then becomes 
essential. One of the most important questions in this context 
is that of self-adjointness of these operators. The two most 
thoroughly studied operators in mathematical physics are the 
Laplacian and the Dirac operator \cite{AsoreyGlobal}.

While a general analysis of the question of self-adjointness of 
these operators on manifolds with boundaries is possible, 
see \cite{AsoreyGlobal, AsoreyEdgeStateStability} for instance, it seems that an investigation of these 
issues in the presence of certain isometries of the underlying 
manifold might be physically useful since, in certain cases, 
the presence of 
these isometries either allows for an exact solution or provides 
sufficient simplification so that most of the relevant aspects of 
the problem can be understood even without having an exact solution. 
This is especially so when one considers systems of relevance in 
physics where often one is interested in understanding the 
qualitative behavior of the system under consideration or where 
the underlying geometry of the problem is simple enough to work out 
the quantitative details. Following this reasoning, in this paper, 
we focus on the properties of the Dirac operator on manifolds 
with boundaries.

It is well known that the question of self-adjointness of operators 
is intimately related to that of the domain on which these 
operators are defined which, in turn, is related to the choice 
of the boundary conditions. The two most commonly used boundary 
conditions are the Dirichlet and the Neumann boundary conditions 
for which either the function or its normal derivative vanishes 
at the boundary respectively. However, for the Laplacian, a more 
general class of boundary condition, the so called Robin boundary 
condition, is also possible. For the Robin boundary condition, 
instead of  requiring the function or its normal derivative to 
vanish, one has the condition that the function and its normal 
derivative are related at the boundary: 
$\psi|_{\partial\mathcal{M}}=\kappa\psi'|_{\partial\mathcal{M}}$.

It is clear that the Dirichlet and the Neumann boundary conditions 
are two extreme cases of the Robin boundary condition corresponding 
to $\kappa=0$ and $\kappa=\infty$, respectively. More interestingly, 
we find that the Robin boundary condition brings in a non-trivial 
length scale in the problem since the parameter $\kappa$ has the 
dimensions of length [L]. Some of the consequences of having this 
non-trivial length parameter were explored in \cite{TRG-Novel} where it 
was shown that Robin boundary condition leads to the presence of 
edge states on the boundary, 
with the number of these edge states being related to $\kappa$ and 
to the geometrical information of the boundary. $\kappa$ plays the role 
of the penetration depth. 

The boundary conditions for the Dirac 
operator were given by the Atiyah, Patodi and Singer \cite{APS1, APS2, APS3}.  
A further generalization of the Atiyah-Patodi-Singer (APS) boundary condition was recently 
given in \cite{BalChiralBags}. As will be seen in more detail subsequently, 
like the Robin boundary condition for the Laplacian, the 
generalized APS boundary condition brings in a length parameter ($\mu$) in the problem. In principle, more general boundary conditions 
are possible but in this paper we will not analyse this issue further since the associated 
isometries are limited.
 
In this paper we focus on the APS boundary condition and 
its generalization with examples where the underlying manifold 
and its boundary have the isometries of sphere. We show that these 
boundary conditions can be thought of as a generalization of the 
Robin boundary condition in the context of the Dirac operator. We 
also show that this boundary condition can lead to several edge 
states on the boundary whose precise number depends on the mass 
parameter $m_{0},\mu$ and the geometry of the boundary. 
We give an analytic expression for the number of edge states as a 
function of these parameters.

In real systems the boundaries are imposed by applying 
suitable potentials. We would like to incorporate this aspect 
even at the mathematical level and as a second important aspect 
of this paper we show how the effect of the boundary conditions 
can be mimicked by replacing the boundary (and with it the 
corresponding boundary condition) by a delta function potential 
at the location of the boundary. In the context of the Schroedinger 
equation in one dimension it is well known that a negative delta 
function potential can lead to the existence of a bound state which 
is generally achieved by demanding the continuity of the wave 
function across the $\delta$-function potential (so that the 
derivative of the wave function is discontinuous). However, as has 
been shown in \cite{BoyaSudarshan}, one can incorporate not only a $\delta$ 
term but also a $\delta'$ term in the potential. Unlike the 
usual case, the analysis proceeds more naturally by considering 
the continuity equation at the location of the 
$\delta, \delta'$ terms. Interestingly, this can be generalized to 
higher dimensions also.

We will show that incorporation of 
the $\delta'$-potential is, in fact, required  
when working with the Dirac operator. As an added advantage of 
replacing the boundary condition by interactions we note that 
since there are no boundaries in the problem any more, the 
question of self-adjointness of the differential operator 
becomes simpler. We show that the role of the parameter $\mu$, 
which in the absence of the boundary condition is not present 
in the problem, is taken up by the strength of the delta function 
potential. We give an exact counting of the number of 
edge states for this problem and show how this is related 
to the number of edge states in the presence of the generalized 
APS boundary condition.

Here we would like to mention some of the recent works incorporating a $\delta$ (and/or a $\delta'$ potential). The effect of singular potentials on vacuum energy has been discussed in \cite{BartonPlasmaShells, GuilarteCastanedaDoubleDelta, ParasharDeltaPlate, CastanedaMosqueraCasimirDelta, CastanedaGuilartedeltadelta'} (including the interaction between extended objects \cite{BordagCastanedaSineGordon}). These papers discuss interesting situations
where the boundary conditions and self-adjointness play
important roles which can possibly be tested experimentally. In addition, the effect of discontinuous potentials on heat kernel expansion has also received some attention \cite{BordagVassilevicHeatKernel, BordagVassilevicNonSmoothQFT} with the focus on self-adjoint Laplacian operators which are positive (since the primary concern was QFTs which are stable). This puts the constraint of looking for boundary conditions satisfying this requirement.
Later we have pointed out that stability can be considered at higher
temperature or with $\phi^4$ interactions (like spontaneous symmetry
breakdown) and an analysis in these circumstances for
heat kernel will complement their analysis.

Finally, to illustrate the significance of the delta function 
potential further, we show that the effect of time dependent 
(or moving) boundaries and boundary conditions can also be mimicked 
by a suitable delta function potential. 
Time dependent boundaries which are ubiquitous in physics, lead to 
time dependent domains in the Hilbert space. In this case the 
meaning of time derivative of the state requires careful analysis.
Replacing the boundary with $\delta$ and $\delta'$ potential enables
physical meaning of these considerations.  
The effect of moving 
boundaries has been investigated briefly in the literature,
see \cite{BerryKlein, MakowskiMovingBoundaries, DembinskiExpandingWell, MarmoMovingWalls, AnzaDilatingDomain}, and 
these are of relevance in diverse physical situations 
like an evolving black hole horizon or the time dependent 
cosmological horizon in an expanding universe, 
the Casimir effect in the presence of 
moving boundaries to name a few.

In short, the objective of this paper is to show that - (i) 
the APS boundary condition can be seen as generalized 
Robin boundary condition and the same method as used to find the number 
of edge states for the Laplacian in the presence of the latter 
\cite{TRG-Novel} can be used to find the number of edge states for the 
Dirac operator in presence of the APS boundary condition, 
(ii) the effect of the Robin or the APS boundary conditions 
(as far as the edge states are concerned) can be mimicked 
by removing the boundary and putting a delta potential of 
suitable strength at the location of the boundary and 
(iii) time varying boundaries can be analyzed by considering 
fixed boundaries but with a time dependent strength of 
the potential.

The paper is organized as follows. In Sec. 2 after briefly 
reviewing the introduction of generalized APS boundary condition, we consider examples of euclidean $\mathbb{R}^2$ and $\mathbb{R}^3$ with a 
two-disc $\mathcal{D}^{2}$ 
and a three-ball $\mathcal{B}^{3}$ removed. Imposing the 
the generalized APS boundary condition we find the exact number of 
edge states as a function of the parameters in the problem. 
Next, in Sec. 3, we show how the effect of the boundary condition 
can be mimicked by introducing a delta function potential of 
suitable strength at the location of the boundary. In Sec. 4 we 
further illustrate the utility of the delta function potential 
by showing how it can be used to study the effects of 
moving boundaries. In Sec. 5 we discuss the possible 
physics applications of these boundary conditions 
(or the corresponding delta function potential). 
We conclude in Sec. 6 with a brief discussion of the results 
and possible future directions.

\section{Edge states for the Dirac operator on flat manifolds with a hole} 

In this section we give a couple of examples of the presence of edge states for the Dirac operator in the presence of boundaries along with analytical expression for their exact number as function of the parameters of the problem. As briefly mentioned before, in the presence of boundaries, the self-adjointness issue for differential operators becomes non-trivial since one has to suitably define the domain of the operator (which includes specification of boundary conditions). For the Dirac operator the self-adjointness analysis in the presence of boundaries leads to the so called APS boundary condition \cite{APS1, APS2, APS3} or its generalization thereof \cite{BalChiralBags}. 

Since our aim is to illustrate the existence of edge states and in finding their exact number (as a function of the parameters in the problem) we consider situations where the system (including the boundary) has the isometries of a sphere. We first consider the problem on $\mathbb{R}^{2}-\mathcal{D}^{2}$, that is, on a two dimensional Euclidean spatial manifold with a disc removed. Next we consider the case of three dimensional spatial manifold with a ball removed ($\mathbb{R}^{3}-\mathcal{B}^{3}$). Interestingly, we will find that whereas there is always at least one edge state in $2+1$ dimensions, this is not the case in $3+1$ dimensions.

\subsection{Dirac operator on $\mathbb{R}^{2}-\mathcal{D}^{2}$}
We are interested in finding the edge states for the Dirac operator on $2+1$ dimensional Minkowski spacetime (with metric signature $+--$) in the presence of a boundary on the spatial manifold in the form of a hole of radius $r_{b}$ (that is, the spatial manifold is $\mathbb{R}^{2}-\mathcal{D}^{2}$). A suitable boundary condition will turn out to be the APS boundary condition \cite{APS1, APS2, APS3}. The presence of edge states for Dirac operator on a disk with APS boundary condition has already been demonstrated in \cite{BalChiralBags} and we will essentially follow the same analysis. In \cite{BalChiralBags} only the existence of edge states was shown and more recently, in \cite{AsoreyBalStability}, a bound on these states was found. Below we give an exact counting of the number of edge states as a function of parameters appearing in the problem.

The Dirac Hamiltonian in one particle quantum mechanics is
\be \label{dirac hamiltonian one particle}
H=-i{\bf \alpha}\cdot\nabla+m_{0}\beta,
\ee
where $\alpha^{i}=\gamma^{0}\gamma^{i}$ and $\beta=\gamma^{0}$ satisfy the condition $(\alpha^{i})^{2}=\beta^{2}=\mathfrak{1}, i=(1,2)$ and we choose $\gamma^{0}=\sigma_{3}$, $\gamma^{i}=i\sigma_{i}$ where $\sigma$'s are the Pauli matrices (note that our convention for the $\gamma$'s is different from that in \cite{BalChiralBags}). 
In polar coordinates the Hamiltonian is
\be \label{dirac hamiltonian 2+1 polar coordinates}
H=-i\sigma_{r}\partial_{r}-\frac{i\sigma_{\theta}}{r}\partial_{\theta}+m_{0}\sigma_{3},
\ee
where $\sigma_{r}=-\sigma_{2}\cos\theta+\sigma_{1}\sin\theta$ and $\sigma_{\theta}=\sigma_{2}\sin\theta+\sigma_{1}\cos\theta$ are explicitly given by

\be \label{sigma}
\sigma_{r}=\left(\begin{array}{cc} 
0 & ie^{-i\theta} \\
-ie^{i\theta} & 0
\end{array}\right), \quad  \quad 
\sigma_{\theta}=\left(\begin{array}{cc} 
0 & e^{-i\theta} \\
e^{i\theta} & 0
\end{array}\right).
\ee

It is easy to check that the total angular momentum operator $J=L+S=-i\partial_{\theta}+\frac{1}{2}\sigma_{3}$ commutes with the Hamiltonian: $[H,J]=0$, implying that the two can be simultaneously diagonalized. The (total) angular momentum eigenstates are: 
\be \label{angular momentum eigen vectors 2+1 dim}
\psi_{j}^{(-)}(\theta)=e^{i(j-\frac{1}{2})\theta}\left(\begin{array}{c} 
1 \\
0
\end{array}\right), \quad \quad \psi_{j}^{(+)}(\theta)=e^{i(j+\frac{1}{2})\theta}\left(\begin{array}{c} 
0 \\
1
\end{array}\right),
\ee

Expressing the eigenvectors $\Psi_{j}(r,\theta)$ of the Hamiltonian \eqref{dirac hamiltonian 2+1 polar coordinates} in terms of the angular momentum eigenvectors \eqref{angular momentum eigen vectors 2+1 dim}: 
\be \label{general hamiltonian eigenvector 2+1 dim}
\Psi_{j}(r,\theta)=p_{j}(r)\psi_{j}^{(+)}(\theta)+q_{j}(r)\psi_{j}^{(-)}(\theta)
\ee
leads to a pair of first order coupled equations for $p_{j}$ and $q_{j}$:
\bea \label{coupled diff eq p' in 2+1 dim}
p'_{j}(r)+\left(j+\frac{1}{2}\right)\frac{p_{j}(r)}{r}+mq_{j}(r) &=& E_{j}q_{j}(r), \\
\label{coupled diff eq q' in 2+1 dim}
-q'_{j}(r)+\left(j-\frac{1}{2}\right)\frac{q_{j}(r)}{r}-mp_{j}(r) &=& E_{j}p_{j}(r).
\eea
These can be decoupled to get the following two second order equations
\be \label{diff eq alpha 2+1 dim}
p''_{j}(r)+\frac{p'_{j}}{r}-\bigg[(m_{0}^{2}-E_{j}^{2})+\frac{\left(j+\frac{1}{2}\right)^{2}}{r^{2}}\bigg]p_{j}(r)=0,
\ee
\be \label{diff eq beta 2+1 dim}
q''_{j}(r)+\frac{q'_{j}(r)}{r}-\bigg[(m_{0}^{2}-E_{j}^{2})+\frac{\left(j-\frac{1}{2}\right)^{2}}{r^{2}}\bigg]q_{j}(r)=0,
\ee
where $E_{j}$ is the eigenvalue corresponding to $\Psi_{j}(r,\theta)$. Equations \eqref{diff eq alpha 2+1 dim} and \eqref{diff eq beta 2+1 dim} fall in the class of Bessel equations. We are interested in edge states or solutions that are localized at the boundary and these are obtained for $\epsilon_{j}^{2}\equiv m_{0}^{2}-E_{j}^{2}>0$ in which case the corresponding independent solutions are given in terms of the modified Bessel functions $I_{j-\frac{1}{2}}(\epsilon_{j} r)$ and $K_{j-\frac{1}{2}}(\epsilon_{j} r)$. Demanding that the solution be square integrable on $(r_{b},\infty)$ leaves us with $K_{j-\frac{1}{2}}(\epsilon_{j} r)$ as the solution for edge states.

An important issue when working on manifolds with boundaries is that of self-adjointness of the Hamiltonian operator and for this we will be following the APS prescription. With the bulk and the boundary inner products defined as $(\Psi,\Phi)=\int_{M}\md V_{M}\Psi^{\dagger}\Phi$ and $\langle\psi,\phi\rangle=\int_{\partial M}\md V_{\partial M}\psi^{\dagger}\phi$, respectively (here $M=\mathbb{R}^{2}-\mathcal{D}^{2}$ and $\Psi|_{\partial M}\equiv\psi$ and $\Phi|_{\partial M}\equiv \phi$ on the boundary $\partial M$ of $M$), we need to evaluate $(\Psi,H\Phi)-(H\Psi,\Phi)$ to find the condition for the self-adjointness of the Hamiltonian \eqref{dirac hamiltonian 2+1 polar coordinates}. 

Imposing single-valuedness of $(\Psi,\Phi)$ and using the fact that these functions fall-off to zero sufficiently fast as $r\rightarrow\infty$ we find that
\be \label{self adjointness eq 2+1 dim}
(\Psi,H\Phi)-(H\Psi,\Phi)=ir_{b}\int\md\theta\Psi^{\dagger}(r_{b},\theta)\sigma_{r}\Phi(r_{b},\theta)=i\langle\psi^{\dagger}(\theta),\sigma_{r}\phi(\theta)\rangle,
\ee
where $\psi(\theta)\equiv\Psi(r_{b},\theta)$ is the value of $\Psi(r,\theta)$ on the boundary at $r=r_{b}$. For $H$ to be self-adjoint, this should be equal to zero.

Following the APS prescription, to make the Hamiltonian into a self-adjoint operator we need to find an operator $K(m_{0})$ on the boundary Hilbert space $\mathcal{H}(\partial\mathcal{M})$ such that it anti-commutes with $\sigma_{r}$ (which appears in the second term of the expression on right in \eqref{self adjointness eq 2+1 dim}), $K(m)\sigma_{r}=-\sigma_{r}K(m)$, such that the eigenvalue of $K^{2}(m_{0})>0$. If such an operator exists then we can split $\mathcal{H}(\partial\mathcal{M})=\mathcal{H}_{1}(\partial\mathcal{M})\oplus\mathcal{H}_{2}(\partial\mathcal{M})$ where $\mathcal{H}_{1}(\partial\mathcal{M})$ and $\mathcal{H}_{2}(\partial\mathcal{M})$ are orthogonal sub-spaces of $\mathcal{H}(\partial\mathcal{M})$.

Because $K$ and $\sigma_{r}$ anticommute, this would imply that for $\psi_{1}(\theta)\in\mathcal{H}_{1}(\partial\mathcal{M})$, $\sigma_{r}\psi_{1}(\theta)\in\mathcal{H}_{2}(\partial\mathcal{M})$. Thus, if we choose the boundary condition so that $\Psi(r_{b},\theta)\equiv\psi(\theta),\Phi(r_{b},\theta)\equiv\phi(\theta)$ are both in the same sub-space, say $\mathcal{H}_{2}(\partial\mathcal{M})$, then the last expression in \eqref{self adjointness eq 2+1 dim} will be equal to zero implying that the Hamiltonian is a symmetric operator (and also self-adjoint, as can be easily checked).

It is easy to check that the operator  
\be \label{consistent boundary operator}
K(m_{0})=-\frac{i}{r_{b}}\left(\sigma_{\theta}\partial_{\theta}-\frac{\sigma_{r}}{2}\right)+m_{0}\sigma_{3}
\ee
satisfies the  first APS requirement in that it anti-commutes with $\sigma_{r}$.

It has recently been suggested in \cite{BalChiralBags} that a generalization of the APS boundary condition is possible and instead of working with the boundary operator $K(m_{0})$ one can also work with a generalized boundary operator defined as
\be 
\bar{K}(\mu)=i\sigma_{r}K(\mu),
\ee
where $K(\mu)$ is as defined in \eqref{consistent boundary operator} with parameter $m_{0}$ replaced by $\mu$.  With this definition the explicit form of the $\bar{K}(\mu)$ operator is 
\be \label{generalized boundary operator 2+1 d} 
\bar{K}(\mu)=\frac{i\sigma_{3}}{r_{b}}\partial_{\theta}-\frac{1}{2r_{b}}+\mu\sigma_{\theta}.
\ee
It can be easily verified that this operator satisfies the first APS requirement $\{\bar{K}(\mu),\sigma_{r}\}=0$. However, note that $K(\mu)$ and $\bar{K}(\mu)$ are related by a similarity transformation and therefore the main message of the generalization of \cite{BalChiralBags} can be taken to be that one can replace the mass parameter $m_{0}$ in \eqref{consistent boundary operator} by a more general parameter $\mu$. In the following we will be working with the operator $\bar{K}(\mu)$ in \eqref{generalized boundary operator 2+1 d}. 

It is also easy to see that $\bar{K}(\mu)$ satisfies the other APS requirement $\bar{K}^{2}(\mu)>0$ as well, with the square of the corresponding eigenvalue given by
\be \label{eigenvalues generalized boundary operator 2+1 d}
\bar{\lambda}^{2}_{j}=\mu^{2}+\frac{j^{2}}{r_{b}^{2}}.
\ee
The corresponding eigenvectors are
\bea \label{positive eigenvector generalized boundary operator 2+1 d}
\psi^{(1)}_{j}(\theta) &=& c_{j}\left[e^{i\left(j+\frac{1}{2}\right)\theta}\left(\begin{array}{c} 
0 \\
1
\end{array}\right)+\frac{1}{\mu}\left(+|\bar{\lambda}_{j}|-\frac{j}{r_{b}}\right)e^{i\left(j-\frac{1}{2}\right)\theta}\left(\begin{array}{c} 
1 \\
0
\end{array}\right)\right], \lambda_{j}>0 \\
\label{negative eigenvector generalized boundary operator 2+1 d}
\psi^{(2)}_{j}(\theta) &=& d_{j}\left[e^{i\left(j+\frac{1}{2}\right)\theta}\left(\begin{array}{c} 
0 \\
1
\end{array}\right)+\frac{1}{\mu}\left(-|\bar{\lambda}_{j}|-\frac{j}{r_{b}}\right)e^{i\left(j-\frac{1}{2}\right)\theta}\left(\begin{array}{c} 
1 \\
0
\end{array}\right)\right], \lambda_{j}<0,
\eea
and the orthogonal sub-spaces $\mathcal{H}^{1}(\partial\mathcal{M})$ and $\mathcal{H}^{2}(\partial\mathcal{M})$ can be identified as those corresponding to positive and negative eigenvalues respectively.

Using the above expressions we now find the boundary conditions on $p_{j}$ and $q_{j}$ by equating the expression for $\Psi_{j}(r,\theta)|_{r=r_{b}}$ as given in \eqref{general hamiltonian eigenvector 2+1 dim} to \eqref{negative eigenvector generalized boundary operator 2+1 d} (which corresponds to $\mathcal{H}^{-}(\partial M)$ for $\bar{K}(\mu)$):
\be \label{boundary condition alpha beta generalized boundary operator 2+1 d}
p_{j}(r_{b})=d_{j}, \quad q_{j}(r_{b})=-\frac{d_{j}}{\mu}\left(|\bar{\lambda}_{j}|+\frac{j}{r_{b}}\right)=-\frac{p_{j}}{\mu}\left(|\bar{\lambda}_{j}|+\frac{j}{r_{b}}\right).
\ee
This can be combined with the following relation between $p_{j}$ and $q_{j}$ (obtained directly from the Dirac equation)
\[
p_{j}(r)=-\frac{q'_{j}(r)}{(E_{j}+m_{0})}+\frac{\left(j-\frac{1}{2}\right)q_{j}(r)}{(E_{j}+m_{0})r},
\]
to get
\be \label{generalized boundary condition 2+1}
\left(|\bar{\lambda}_{j}|+\frac{j}{r_{b}}\right)q'_{j}(r_{b})=\left[\mu(E_{j}+m_{0})+\frac{1}{r_{b}}\left(j-\frac{1}{2}\right)\left(|\bar{\lambda}_{j}|+\frac{j}{r_{b}}\right)\right]q_{j}(r_{b}).
\ee

The above boundary condition can be thought of as the generalization of the Robin boundary condition $\psi(r_{b})=\kappa\psi'(r_{b})$ with $\kappa$ in the present case being a function of the parameters $m_{0}$, $\mu$ and $r_{b}$ as well as a function of the spin $j$. After using the explicit form for $q_{j}(r)=cK_{j-\frac{1}{2}}(\epsilon_{j} r)$ (where $c$ is a constant of integration) this can be written as
\be \label{boundary condition beta generalized boundary operator 2+1 d}
\mu(E_{j}+m_{0})K_{j-\frac{1}{2}}(\epsilon_{j} r_{b})+\epsilon_{j}\left(|\bar{\lambda}_{j}|+\frac{j}{r_{b}}\right)K_{j+\frac{1}{2}}(\epsilon_{j} r_{b})=0,
\ee
where, in writing the above expression we used the identity $K_{\nu}'(x)=\frac{\nu K_{\nu}(x)}{x}-K_{\nu+1}(x)$.

\section*{Counting the number of edge states}
We now count the number of edge states corresponding to the above boundary condition. We start by noting that since $(E_{j}+m_{0})$, $\epsilon_{j}$ and $(|\bar{\lambda}_{j}|+j/r_{b})$ are all greater than zero, \eqref{boundary condition beta generalized boundary operator 2+1 d} can be satisfied only for $\mu<0$. In other words, there will be edge state solutions only for $\mu<0$. 

In \eqref{boundary condition beta generalized boundary operator 2+1 d} there appear three parameters $(\mu,m_{0},r_{b})$ and it would seem that to get an exact analytic dependence of the number of edge states on all the three parameters might be difficult. However, by redefining $(\mu,m_{0})$ in terms of $r_{b}$ we can eliminate the radius $r_{b}$ of the hole from the problem. Specifically, we define 
\[
\mu=\frac{a}{r_{b}}, \quad \quad \quad m_{0}=\frac{b}{r_{b}}.
\]
With this redefinition it is easy to verify that \eqref{boundary condition beta generalized boundary operator 2+1 d} can be written as 
\be
a(\pm\sqrt{b^{2}-\epsilon_{j}^{2}r_{b}^{2}}+b)K_{j-\frac{1}{2}}(\epsilon_{j} r_{b})+\epsilon_{j} r_{b}(\sqrt{a^{2}+j^{2}}+j)K_{j+\frac{1}{2}}(\epsilon_{j} r_{b})=0,
\ee
or writing $\epsilon_{j} r_{b}\equiv x$ and $a=-A$, $A>0$ (recall that it was noted above that $\mu<0$ to obtain edge states)  we can rewrite the above equation as
\be \label{generalized boundary operator 2+1 redefined}
A=\frac{x(\sqrt{A^{2}+j^{2}}+j)K_{j+\frac{1}{2}}(x)}{(\pm\sqrt{b^{2}-x^{2}}+b)K_{j-\frac{1}{2}}(x)}.
\ee
In the above equation the $\pm$ sign in the denominator will correspond to $j>0$ and $j<0$ respectively.

Next we note that the $x$-derivative of the rhs of the above equation in the limit $x\rightarrow0$ is zero (the only requirement for this argument to work is that $j>1/2$ and $j<-1/2$ for the two cases mentioned above). Since the tangent to the rhs is zero as a function of $x$ in the limit $x\rightarrow0$ and that of lhs is obviously so ($A$ being independent of $x$), the maximum number of bound state solutions will be given by finding the first curve (corresponding to the rhs of \eqref{generalized boundary operator 2+1 redefined}) which does not intersect the $A=\mathrm{const.}$ line in the limit $x\rightarrow0$ (since all the curves corresponding to the rhs for different values of $j$ start out horizontally by the above argument).

We first consider the case $j>0$ which corresponds to choosing the plus sign in the denominator of \eqref{generalized boundary operator 2+1 redefined} (and as already mentioned, we need $j>1/2$ for the argument to  
work). For $x\rightarrow0$ the modified Bessel function can be approximated by
\be \label{approximate K}
\lim_{x\rightarrow0} K_{\nu}(x)\approx\frac{\Gamma(\nu)}{2}\left(\frac{2}{x}\right)^{\nu}, \sqrt{\nu+1}\gg x>0.
\ee
Using this on the rhs of \eqref{generalized boundary operator 2+1 redefined} and remembering that $j$ is a positive odd-half integer and that $\Gamma(n)=(n-1)!$ we find that in the limit $x\rightarrow0$ the condition in \eqref{generalized boundary operator 2+1 redefined} implies
\be \label{equation for number of edge states plus 2+1}
Ab=\left(j-\frac{1}{2}\right)(j+\sqrt{A^{2}+j^{2}}).
\ee
The maximum number of allowed solutions $j_{\mathrm{max}}$ will be given in terms of the solution of this  equation
\be \label{maximum number of edge states plus 2+1} 
j_{\mathrm{max}}=\frac{A+b+b\sqrt{1+4A^{2}+8Ab}}{2(A+2b)}.
\ee
The other solution of \eqref{equation for number of edge states plus 2+1} gives $j<0$ and hence does not apply.

\begin{figure}[htbp]
\begin{center}
\includegraphics[scale=1]{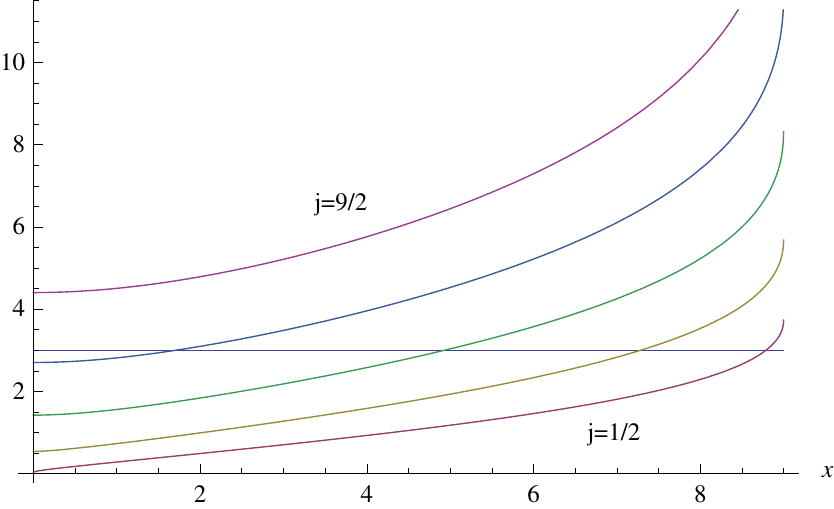}
\end{center}
\caption{\label{edge states for positive j}} Solution for the boundary condition \eqref{generalized boundary operator 2+1 redefined} in the presence of generalized boundary operator in two dimensions for $A=3, b=9, j>0$.
\end{figure}

Next we consider the case $j<0$ corresponding to choosing the minus sign in the denominator of \eqref{generalized boundary operator 2+1 redefined} (and as mentioned earlier, we need $j<-1/2$ for the argument to work). 
Since $j<-1/2$ we cannot use the approximation in \eqref{approximate K}. However, the modified Bessel functions satisfy the relation $K_{\nu}(x)=K_{-\nu}(x)$ and therefore $K_{j+\frac{1}{2}}(x)=K_{-j-\frac{1}{2}}(x)$ and $K_{j-\frac{1}{2}}(x)=K_{-j+\frac{1}{2}}(x)$. With these forms the condition in the approximation in \eqref{approximate K} applies and we can now proceed to evaluate the rhs of the above equation in the limit $x\rightarrow0$. Here we also need to be careful about the factor in the denominator: $b-(b^{2}-x^{2})^{1/2}\approx x^{2}/2b$. With all these we finally get the equation 
\be \label{equation for number of edge states minus 2+1}
-\left(j+\frac{1}{2}\right)A=b(j+\sqrt{A^{2}+j^{2}}).
\ee
The solution of this equation which gives the minimum value of $j$ (remember $j<0$) that still leads to an edge state is
\be \label{minimum number of edge states minus 2+1} 
j_{\mathrm{min}}=-\frac{A+b+b\sqrt{1+4A^{2}+8Ab}}{2(A+2b)},
\ee
which just differs by an overall sign from the corresponding expression in \eqref{maximum number of edge states plus 2+1} for $j>0$. 

One can easily convince oneself that for $j>0$ and when $j_{\mathrm{max}}$ is not an integer, the actual number of edge states is given by the lowest integer greater than $j_{\mathrm{max}}$ in \eqref{maximum number of edge states plus 2+1}. Similarly for $j<0$ and when $j_{\mathrm{min}}$ is not an integer, the actual number of edge states is the greatest integer smaller than $j_{\mathrm{min}}$ in \eqref{minimum number of edge states minus 2+1}. When $j_{\mathrm{max}}$ and $j_{\mathrm{min}}$ are integers than the number of edge states is $j_{\mathrm{max}}-1$ and $j_{\mathrm{min}}+1$ since only for these values of $j$ does the rhs of \eqref{generalized boundary operator 2+1 redefined} intersects the $A=\mathrm{const}$ curve (the curves corresponding to $j_{\mathrm{max}}$ and $j_{\mathrm{min}}$ are coincident with the $A=\mathrm{const}$ curve at $x=0$ implying that there is no intersection).

\begin{figure}[htbp]
\begin{center}
\includegraphics[scale=1]{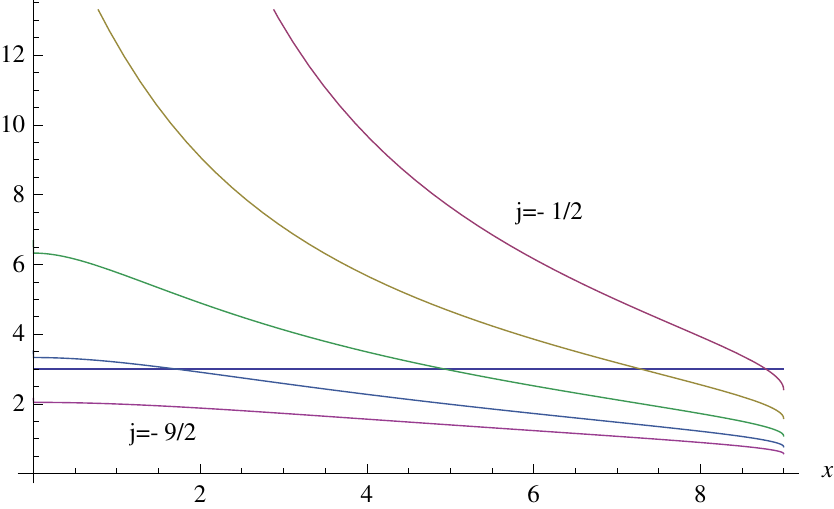}
\end{center}
\caption{\label{edge states for negative j}} Solution for the boundary condition \eqref{generalized boundary operator 2+1 redefined} in the presence of generalized boundary operator in two dimensions for $A=3, b=9, j<0$.
\end{figure}

The correctness of the above counting (including the arguments leading upto it) can also be seen in Figs. \ref{edge states for positive j} and \ref{edge states for negative j} corresponding to $j>0$ and $j<0$, respectively, for the parameter choice $A=3$ and $b=9$. From the figures we see that there are four edge states for each case. Our formulas \eqref{maximum number of edge states plus 2+1} and \eqref{minimum number of edge states minus 2+1} give $j_{\mathrm{max}}=-j_{\mathrm{min}}=3.69$ and therefore our counting prescription implies that the number of edge states is $4+4$ in agreement with what is seen in the figures.

\subsection*{Number of edge states in various limits in $2+1$ dimensions}

It might be of some interest to see the behavior of the number of edge states when the parameter $A$ in \eqref{maximum number of edge states plus 2+1} takes on certain limiting values (we only focus on $j_{\mathrm{max}}$, the case of $j_{\mathrm{min}}$ being identical). It is easy to see that 
\begin{itemize}
\item
In the limit $A\rightarrow0$, $j_{\mathrm{max}}\rightarrow\frac{1}{2}$; that is, their is only a single edge state.
\item
In the limit $A\rightarrow\infty$, $j_{\mathrm{max}}\rightarrow b+\frac{1}{2}$; that is, the number of edge states grows linearly with the mass of the Dirac particle when $A$ is very large.
\end{itemize} 

\subsection{Dirac operator on $\mathbb{R}^{3}-\mathcal{B}^{3}$}

We next consider the Dirac operator in $3+1$ dimensional Minkowski spacetime (with metric signature $+---$) with a three ball of radius $r_{b}$ removed from the the spatial manifold, that is, we work on the spatial slice $\mathbb{R}^{3}-\mathcal{B}^{3}$. Since the analysis of the Dirac equation in presence of spherical symmetry is standard, in the following we will be brief. The Dirac Hamiltonian is 
\be \label{dirac hamiltonian 3+1}
H=-i\alpha\cdot\nabla+m_{0}\beta,
\ee
with $\alpha^{i}=\gamma^{0}\gamma^{i}$ and $\beta=\gamma^{0}$. We work in the representation  
\be \label{representation for gamma matrices 3+1}
\alpha^{i}=\left(\begin{array}{cc} 
0 & \sigma_{i} \\
\sigma_{i} & 0
\end{array}\right), \quad  \quad 
\beta=\left(\begin{array}{cc} 
1 & 0 \\
0 & -1
\end{array}\right).
\ee
Using the spherical polar coordinates the Hamiltonian can be written as
\be \label{dirac hamiltonian spherical coord 3+1}
H=-i\gamma^{r}-\frac{i\gamma^{\theta}}{r}\partial_{\theta}-\frac{i\gamma^{\varphi}}{r}\partial_{\varphi}+m_{0}\gamma^{0}.
\ee
In the above equation
\be 
\gamma^{r}=\left(\begin{array}{cc} 
0 & \sigma_{r} \\
\sigma_{r} & 0
\end{array}\right), \quad  \quad 
\gamma^{\theta}=\left(\begin{array}{cc} 
0 & \sigma_{\theta} \\
\sigma_{\theta} & 0
\end{array}\right), \quad \quad 
\gamma^{\varphi}=\left(\begin{array}{cc} 
0 & \sigma_{\varphi} \\
\sigma_{\varphi} & 0
\end{array}\right)
\ee
with
\be
\sigma_{r}=\left(\begin{array}{cc} 
\cos\theta & \sin\theta e^{-i\varphi} \\
\sin\theta e^{i\varphi} & -\cos\theta
\end{array}\right), \quad 
\sigma_{\theta}=\left(\begin{array}{cc} 
-\sin\theta & \cos\theta e^{-i\varphi} \\
\cos\theta e^{i\varphi} & \sin\theta
\end{array}\right), \quad 
\sigma_{\varphi}=\left(\begin{array}{cc} 
0 &  \frac{-i}{\sin\theta}e^{-i\varphi} \\
\frac{i}{\sin\theta}e^{i\varphi} & 0
\end{array}\right).
\ee

Because of spherical symmetry the Hamiltonian commutes with the square of the total angular momentum $J^{2}$, with $\overrightarrow{J}=\overrightarrow{L}+\overrightarrow{S}$ ($\overrightarrow{L}$ being the orbital angular momentum and $\overrightarrow{S}$ the spin), as well as with its $z$-component $J_{z}$, implying that these three operators can be diagonalized simultaneously. Following \cite{Merzbacher}, the eigenvectors for the Hamiltonian can be written as
\be \label{eigenvector hamiltonian 3+1}
\Psi_{1}(r,\theta,\varphi)=\left(\begin{array}{c} 
F(r)\mathcal{Y}^{jm}_{j-\frac{1}{2}} \\
-if(r)\mathcal{Y}^{jm}_{j+\frac{1}{2}}
\end{array}\right), \quad \quad \Psi_{2}(r,\theta,\varphi)=\left(\begin{array}{c} 
G(r)\mathcal{Y}^{jm}_{j+\frac{1}{2}} \\
-ig(r)\mathcal{Y}^{jm}_{j-\frac{1}{2}}
\end{array}\right),
\ee
where
\[
\mathcal{Y}^{jm}_{l}=\mathcal{Y}^{l\pm\frac{1}{2},m}_{l}=\frac{1}{\sqrt{2l+1}}\left(\begin{array}{c} 
\pm\sqrt{l\pm m+\frac{1}{2}}Y^{m-\frac{1}{2}}_{l} \\
\sqrt{l\mp m+\frac{1}{2}}Y^{m+\frac{1}{2}}_{l}
\end{array}\right).
\]
(One can check that $\mathcal{Y}^{jm}_{j-\frac{1}{2}}$ and $\mathcal{Y}^{jm}_{j+\frac{1}{2}}$ are eigenstates of $J^{2}$.)

Since we are interested in studying the effects of boundaries and boundary conditions, in the following we will concentrate only on $\Psi_{1}(r,\theta,\varphi)$, the analysis for $\Psi_{2}(r,\theta,\varphi)$ being analogous.The radial equations for $F(r)$ and $f(r)$ resulting from $H\Psi(r,\theta,\varphi)=E\Psi(r,\theta,\varphi)$ are:
\bea \label{radial equations 3+1}
(E-m_{0})F-\left[\frac{\md f}{\md r}+\left(j+\frac{3}{2}\right)\frac{f(r)}{r}\right] &=& 0, \\
(E+m_{0})f+\left[\frac{\md F}{\md r}-\left(j-\frac{1}{2}\right)\frac{F(r)}{r}\right] &=& 0. 
\eea
Eliminating $f(r)$ in favor of $F(r)$ and defining $n=j-1/2$, $m^{2}_{0}-E^{2}=\epsilon_{j}^{2}$ and $\epsilon_{j} r=x$ we get the equation
\be \label{second order radial eq 3+1}
(\epsilon_{j} r)^{2}\frac{\md^{2}F(r)}{\md (\epsilon_{j} r)^{2}}+2\epsilon_{j} r\frac{\md F(r)}{\md (\epsilon_{j} r)}-[(\epsilon_{j} r)^{2}+n(n+1)]F(r)=0.
\ee
This equation has two solutions $c_{1}I_{n+\frac{1}{2}}(\epsilon_{j} r)/r^{1/2}$ and $c_{2}K_{n+\frac{1}{2}}(\epsilon_{j} r)/r^{1/2}$ of which only the second solution is finite on $r_{b}\leq r\le\infty$ (here $c_{1}$ and $c_{2}$ are arbitrary constants).

Following an analysis similar to that of the previous sub-section we find that the boundary operator which makes the Hamiltonian a self-adjoint operator is 
\be \label{boundary operator 3+1}
K(m_{0})=\frac{i\gamma^{r}}{r_{b}}-\frac{i\gamma^{\theta}}{r_{b}}\partial_{\theta}-\frac{i\gamma^{\varphi}}{r_{b}}\partial_{\varphi}+m_{0}\gamma^{0},
\ee
or its generalization in the form $\bar{K}(\mu)=i\gamma^{r}K(\mu)$ (which is again related to $K(m_{0})$ by a similarity transformation)  
\be \label{generalized boundary operator 3+1}
\bar{K}(\mu)=\frac{\bar{\gamma}^{\theta}}{r_{b}}\partial_{\theta}+\frac{\bar{\gamma}^{\varphi}}{r_{b}}\partial_{\varphi}+i\mu\bar{\gamma}^{0}-\frac{1}{r_{b}},
\ee
where $\bar{\gamma}^{\theta}=\gamma^{r}\gamma^{\theta}$ etc. We choose to work with the operator in \eqref{generalized boundary operator 3+1} and find the boundary condition resulting from it.

The eigenvectors of $\bar{K}(\mu)$ are
\bea \label{eigenvectors generalized boundary operator 3+1}
\psi_{j}^{+}(\theta,\phi) &=& c_{j}\bigg[\frac{1}{\mu}\left(|\lambda_{j}|-\frac{1}{r_{b}}\left(j+\frac{1}{2}\right)\right)\left(\begin{array}{c} 
\mathcal{Y}^{jm}_{j-\frac{1}{2}} \\
0
\end{array}\right)-i\left(\begin{array}{c} 
0 \\
\mathcal{Y}^{jm}_{j+\frac{1}{2}}
\end{array}\right)\bigg], \\
\psi_{j}^{-}(\theta,\phi) &=& d_{j}\bigg[-\frac{1}{\mu}\left(|\lambda_{j}|+\frac{1}{r_{b}}\left(j+\frac{1}{2}\right)\right)\left(\begin{array}{c} 
\mathcal{Y}^{jm}_{j-\frac{1}{2}} \\
0
\end{array}\right)-i\left(\begin{array}{c} 
0 \\
\mathcal{Y}^{jm}_{j+\frac{1}{2}}
\end{array}\right)\bigg],
\eea
with the corresponding eigenvalues
\be \label{eigenvalue generalized boundary operator 3+1}
\lambda_{j}=\pm\sqrt{\mu^{2}+\frac{1}{r_{b}^{2}}\left(j+\frac{1}{2}\right)^{2}}.
\ee

As for the $2+1$ dimensional case, if we now demand that $\Psi(r_{b},\theta,\varphi)=\psi_{j}^{-}(\theta,\varphi)$ we end up with the boundary condition
\be \label{boundary condition generalized boundary operator 3+1}
\left(|\lambda_{j}|+\frac{1}{r_{b}}\left(j+\frac{1}{2}\right)\right)F'(r_{b})=\left[\mu(E+m_{0})+\frac{1}{r_{b}}\left(j-\frac{1}{2}\right)\left(|\lambda_{j}|+\frac{1}{r_{b}}\left(j+\frac{1}{2}\right)\right)\right]F(r_{b}).
\ee
From \eqref{second order radial eq 3+1} we know that $F(r)=cK_{j}(\epsilon_{j} r)/r^{1/2}$. To simplify the subsequent expression(s) we now express $\mu$ and $m_{0}$ in terms of $r_{b}$ as $\mu=a/r_{b}$ and $m_{0}=b/r_{b}$ and define $x=\epsilon_{j}r_{b}$. With all this the equation for the boundary condition becomes
\be \label{edge state equation generalized 3+1}
a=\frac{-x\left[\sqrt{a^{2}+\left(j+\frac{1}{2}\right)^{2}}+\left(j+\frac{1}{2}\right)\right]}{(\sqrt{b^{2}-x^{2}}+b)}\frac{K_{j+1}(x)}{K_{j}(x)}.
\ee

As in $2+1$ dimensions it turns out that we get non-trivial solutions only for $a<0$. Making the replacement $a\rightarrow-A$ and comparing the resulting equation with the corresponding equation \eqref{generalized boundary operator 2+1 redefined} for $2+1$ dimensions we find that the present equation can be obtained by replacing $j$ with $j+1/2$ in \eqref{generalized boundary operator 2+1 redefined}. All the previous arguments therefore go through and the number of edge solutions can also be obtained by replacing $j_{\mathrm{max}}$ in \eqref{maximum number of edge states plus 2+1} with $j_{\mathrm{max}}+1/2$ which gives the maximum number of edge states as 
\be \label{maximum number of edge states generalized 3+1}
j_{\mathrm{max}}=\frac{-b+b\sqrt{1+4A^{2}+8Ab}}{2(A+2b)},
\ee

\begin{figure}[htbp]
\begin{center}
\includegraphics[scale=0.7]{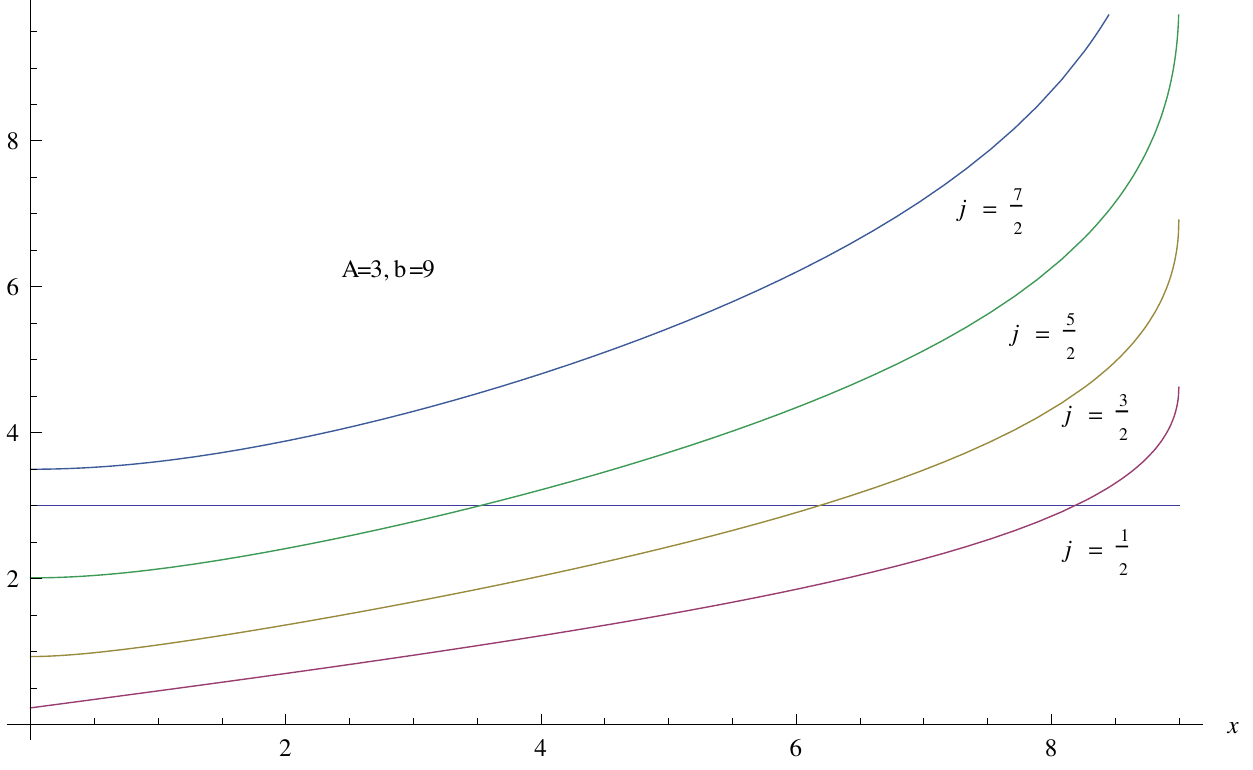}
\end{center}
\caption{\label{edge states for generalized boundary operator in $3+1$ dimensions}} Solution for the boundary condition \eqref{edge state equation generalized 3+1} in the presence of generalized boundary operator in three spatial dimensions for $A=3, b=9$.
\end{figure}

As an example consider $A=3, b=9$ for which the above equation gives $j_{\mathrm{max}}=3.1941$. Graphically the situation is shown in figure \ref{edge states for generalized boundary operator in $3+1$ dimensions}. If we write $j_\mathrm{max}=n_{m}+f$, where $n_{m}\in\mathbb{Z}$ and $f$ is the fractional part, then for $f<1/2$, the number of edge states is $n_{m}$ and for $f>1/2$ the number of edge states is $n_{m}+1$.  

\subsection*{Number of edge states in various limits in $3+1$ dimensions}
As for $2+1$ dimensions, we can calculate the behavior of the number of edge states in various limits of the parameter $A$ in \eqref{maximum number of edge states generalized 3+1} for $3+1$ dimensions. We find that
\begin{itemize}
\item
In the limit $A\rightarrow0$, $j_{\mathrm{max}}\rightarrow0$. This is to be contrasted with the $2+1$ dimensional case where one edge state survives in this limit.
\item
In the limit $A\rightarrow\infty$, $j_{\mathrm{max}}\rightarrow b$. This is similar to $2+1$ dimensions where the behavior in this limit is $b+1/2$.
\end{itemize}

\section{Singular potentials and the boundary effects}

In this section we suggest an interesting possibility whereby the 
effect of the boundary and the corresponding boundary condition 
can be mimicked by replacing the boundary and the boundary 
condition by delta function potential with an appropriately 
chosen strength at the location of the original boundary. In essence, 
what is being suggested is that if the original domain for the 
operator was restricted to the interval $r_{b}\leq r<\infty$ with 
suitable boundary condition at $r=r_{b}$ then one can extend the domain 
to the interval $0\leq r<\infty$ (if we think of $r$ as the radial 
coordinate) and introduce a delta function potential $P\delta(r-r_{b})$ 
in the Hamiltonian with a suitably chosen value for $P$. This has 
the additional advantage that in the absence of a boundary it becomes 
easier to address the self-adjointness issue. From physics 
perspective also this makes sense since, in general, boundaries (and boundary conditions) 
are achieved by introducing suitable potentials in the 
experimental set up.

To set things in context we first recall the text book example 
involving the Laplacian in one dimension where, in presence of the 
delta function potential one can get a single bound state. We will 
show that in higher dimensions the number of bound states can depend 
on the strength of the delta potential. As has been shown in 
\cite{BoyaSudarshan} it is not necessary to restrict to only 
$\delta$ potential and one can also include its derivative.
The key point is that instead of working with the Schroedinger equation, one uses the conservation of the probability current (in other words, one uses the continuity equation). 
Interestingly, while the $\delta$ function allows for the discontinuity
in the derivative of the wave function the $\delta'$ potential allows 
for the discontinuity in wave function itself. But the probability
remains continuous.  
Although in the context of the Laplacian the inclusion of the 
$\delta'$ term might appear a bit adhoc, we will see that 
for the Dirac operator the $\delta'$ term arises naturally.

To have a concrete quantitative understanding of what is 
being proposed we first consider the 
case of the Laplacian on $\mathbb{R}^{2}-\mathcal{D}^{2}$. This problem 
was considered in \cite{TRG-Novel} where an exact expression 
was obtained for the number of bound (edge) states in the 
presence of the Robin boundary condition 
$\kappa\psi+\partial_{\overrightarrow{n}}\psi=0$. 
We now consider the problem where the disc $\mathcal{D}^{2}$ is 
restored and the Robin boundary condition is replaced by a 
delta function potential $V(r)=P\delta(r-r_{b})$ where 
$P$ is the strength of the potential.

The Schroedinger equation takes the form 
\be \label{laplacian in 2+1}
\frac{\partial^{2}\psi}{\partial r^{2}}+\frac{1}{r^{2}}\frac{\partial^{2}\psi}{\partial\theta^{2}}+\frac{1}{r}\frac{\partial\psi}{\partial r}+P\delta(r-r_{b})\psi=\lambda\psi.
\ee
For $r\neq r_{b}$ this equation is solved by
\bea 
\psi(r<r_{b},\theta) &=& cI_{n}(\sqrt{\lambda}r)e^{in\theta}, \\ 
\psi(r>r_{b},\theta) &=& dK_{n}(\sqrt{\lambda}r)e^{in\theta}.
\eea
At $r=r_{b}$ we need to worry about the delta function singularity 
and we integrate the radial equation (obtained after using 
$\psi=R(r)e^{in\theta}$ in \eqref{laplacian in 2+1}) from 
$r_{b}-\epsilon$ to $r_{b}+\epsilon$ and take the limit 
$\epsilon\rightarrow0$. This gives the equation
\be \label{discontinuity in the derivative for laplacian}
\left[\frac{\md R}{\md r}\right]_{r_{b}-\epsilon}^{r_{b}+\epsilon}+PR(r_{b})=0.
\ee
In writing this equation we have assumed the continuity of the 
wave function across $r=r_{b}$ which implies 
$cI_{n}(\sqrt{\lambda}r_{b})=dK_{n}(\sqrt{\lambda}r_{b})$. 
Using 
$I'_{\nu}(x)=I_{\nu-1}(x)-\nu I_{\nu}(x)/x$ and $K'_{\nu}(x)=-K_{\nu-1}(x)-\nu K_{\nu}(x)/x$ 
(where prime implies derivative with respect to the argument) 
then gives the condition
\be 
x\left[\frac{K_{n-1}(x)}{K_{n}(x)}+\frac{I_{n-1}(x)}{I_{n}(x)}\right]=Q,
\ee
where we have defined $P=Q/r_{b}$ and $x=\sqrt{\lambda}r_{b}$.

Now it turns out that the tangent to the curves corresponding to 
the expression on the left of the above equation is horizontal 
near $x=0$ and therefore the arguments used in the previous 
section to find analytic expression for the number of edgestates 
can be applied. Using the approximation in \eqref{approximate K} 
(and a similar approximation for $I_{\nu}(x)$) we find that the 
above equation in the limit $x\rightarrow0$ leads to the 
following expression for the maximum number of bound states:
\be \label{maximum number of bound states delta potential laplacian 2+1}
\lim_{x\rightarrow0}x\left(\frac{x}{2(n_{m}-1)}+\frac{2n_{m}}{x}\right)=2n_{m}=Q.
\ee

On the other hand, in \cite{TRG-Novel}, in presence of the 
Robin boundary condition the maximum number of edge states was 
found to be $n_{m}=-\kappa r_{b}$ ($\kappa<0$). Comparing this 
with the previous expression we find that the problem in the 
presence of delta potential but with no boundaries is equivalent 
to the problem in the presence of Robin boundary condition 
if we choose $Q=-2\kappa r_{b}$ or equivalently
\be \label{equality of delta potential and robin boundary condition laplacian 2+1}
P=-2\kappa.
\ee

Having established the correspondence between Robin boundary 
condition and the presence of a delta function potential 
(of appropriate strength) at the location of the boundary for 
the Laplacian, the next task would be to do the same for the 
Dirac operator with APS boundary condition. However, before doing 
that we would like to point to the analysis by Boya and 
Sudarshan \cite{BoyaSudarshan} where it is shown that the 
usual matching conditions employed in problems involving delta 
potential can be seen as arising naturally from the current 
conservation equation (the continuity equation) for the Schroedinger 
equation. Additionally, they show that use of current 
conservation as the basic principle allows for more general 
singular potentials - those involving both $\delta(r)$ and 
$\delta'(r)$ type singularity - and these can be handled by 
essentially similar methods. Inclusion of $\delta'(r)$ 
type potential directly at the level of the Schroedinger equation, 
requires careful analysis.

From the point of view of the Laplacian or the Schroedinger equation, inclusion of a $\delta'(r)$ potential might seem a bit adhoc. However, we next show that when considering the Dirac operator the $\delta'(r)$ type term in the potential arises naturally. We begin by considering the first example of the previous section but now with the hole of radius $r_{b}$ replaced by a delta function potential $V(r)=-P\delta(r-r_{b})$. The Hamiltonian is therefore the same as in \eqref{dirac hamiltonian 2+1 polar coordinates} except for the addition of $V(r)$:
\be \label{dirac hamiltonian 2+1 polar coordinates delta potential}
H=-i\sigma_{r}\partial_{r}-\frac{i\sigma_{\theta}}{r}\partial_{\theta}+m\sigma_{3}+V(r).
\ee

As usual, we want to solve the equation $H\psi_{j}(r,\theta)=E_{j}\psi_{j}(r,\theta)$ and since the system is still spherically symmetric, the wavefunction is still expressed as in \eqref{general hamiltonian eigenvector 2+1 dim}. In the presence of a potential $V(r)=P\delta(r-r_{b})$ the first order Dirac equations \eqref{coupled diff eq p' in 2+1 dim} and \eqref{coupled diff eq q' in 2+1 dim} are modified by the addition of the potential term to
\bea \label{general diff eq 2+1 dim delta potential 1}
p'_{j}(r)+\left(j+\frac{1}{2}\right)\frac{p_{j}(r)}{r}+mq_{j}(r)+V(r)q_{j}(r) &=& E_{j}q_{j}(r), \\
\label{general diff eq 2+1 dim delta potential 2}
-q'_{j}(r)+\left(j-\frac{1}{2}\right)\frac{q_{j}(r)}{r}-mp_{j}(r)+V(r)p_{j}(r) &=& E_{j}p_{j}(r).
\eea
These equations can be decoupled as 
before and, for instance, leads to the following second order equation for $q_{j}(r)$ (compare with \eqref{diff eq beta 2+1 dim} corresponding to $V(r)=0$) 
\bea \label{diff eq q 2+1 dim with delta potential}
&&q''_{j}(r)+\frac{q'_{j}(r)}{r}-\bigg[(m^{2}-(E_{j}-P\delta(r-r_{b}))^{2})
+\frac{\left(j-\frac{1}{2}\right)^{2}}{r^{2}}\bigg]q_{j}(r)\nonumber\\
&&-\left[rq'_{j}(r)-\left(j-\frac{1}{2}\right)q_{j}(r)\right]
\frac{P\delta'(r-r_{b})}{r(E_{j}-V(r)+m)}=0.
\eea
Since the operator above is essentially the Laplacian, we see that 
inclusion of the delta potential in the Dirac operator automatically 
leads to the presence of $\delta'(r)$ terms for the Laplacian. 

We can now go on to show that even for the Dirac operator the effect of the APS boundary condition, as discussed in the previous section, can be mimicked by the delta potential. In other words, as for the Laplacian, we can choose the strength $P$ of the delta potential appropriately so that the number of edge states for the two problems are identical. For $r<r_{b}$ and for $r>r_{b}$, $V(r)=0$ and, therefore, in these regions the second order equations for $p_{j}$ and $q_{j}$ will be the same as in \eqref{diff eq alpha 2+1 dim} and \eqref{diff eq beta 2+1 dim} with the solutions given in terms of the modified Bessel functions. Specifically, if we choose the $q_{j}(r)$ solution as 
\be \label{q solution 2+1}
q_{j}(r<r_{b})=c_{1}I_{j-\frac{1}{2}}(\epsilon_{j} r), \quad q_{j}(r>r_{b})=c_{2}K_{j-\frac{1}{2}}(\epsilon_{j} r),
\ee
then the corresponding solutions for $p_{j}(r)$ (using \eqref{coupled diff eq q' in 2+1 dim}) are 
\be \label{p solution 2+1}
p_{j}(r<r_{b})=-\frac{c_{1}\epsilon_{j}}{m_{0}\pm\sqrt{m_{0}^{2}-\epsilon_{j}^{2}}}I_{j+\frac{1}{2}}(\epsilon_{j} r), \quad p_{j}(r>r_{b})=\frac{c_{2}\epsilon_{j}}{m_{0}\pm\sqrt{m_{0}^{2}-\epsilon_{j}^{2}}}K_{j+\frac{1}{2}}(\epsilon_{j} r),
\ee
where $\epsilon_{j}=\sqrt{m_{0}^{2}-E_{j}^{2}}$.

As for the Laplacian (Schroedinger equation), we might try and integrate \eqref{diff eq q 2+1 dim with delta potential}. However, the presence of the $\delta'$ terms does not allow a straight forward use of this standard procedure. One might, presumably, generalize the method of Boya and Sudarshan \cite{BoyaSudarshan} and use the current conservation equation for the Dirac Hamiltonian. However, in this paper we follow the procedure of \cite{LoeweDiracDelta, BenguriaDiracDelta} and directly work with the first order equations \eqref{general diff eq 2+1 dim delta potential 1} and \eqref{general diff eq 2+1 dim delta potential 2}.

We multiply \eqref{general diff eq 2+1 dim delta potential 1} by $p_{j}$ and \eqref{general diff eq 2+1 dim delta potential 2} by $q_{j}$ and add the two equations to obtain:
\be 
\frac{(p_{j}^{2})'}{2}+\frac{(q_{j}^{2})'}{2}=-\frac{j(p_{j}^{2}-q_{j}^{2})}{r}-\frac{(p_{j}^{2}+q_{j}^{2})}{2r}-2mp_{j}q_{j}.
\ee
We integrate the above equation from $r_{b}-\epsilon$ to $r_{b}+\epsilon$ and take the limit $\epsilon\rightarrow0$ to the condition
\be \label{effect of delta potential 1}
p_{j>}^{2}+q_{j>}^{2}=p_{j<}^{2}+q_{j<}^{2},
\ee
where $p_{j\lessgtr}=p_{j}(r_{b}\mp\epsilon)$ and $q_{j\lessgtr}=q_{j}(r_{b}\mp\epsilon)$. This condition simply fixes the normalization of the solution on one side $r=r_{b}$ in terms of the normalization chosen on the other side.

Next we multiply \eqref{general diff eq 2+1 dim delta potential 1} by $q_{j}$ and \eqref{general diff eq 2+1 dim delta potential 2} by $p_{j}$ and subtract the two equations. This gives:
\be
p_{j}q'_{j}-p'_{j}q_{j}=\frac{2jp_{j}q_{j}}{r}-(E_{j}-m)q_{j}^{2}-(E_{j}+m)p_{j}^{2}-P\delta(r-r_{b})(p_{j}^{2}+q_{j}^{2}).
\ee
We divide the above equation by $p_{j}^{2}+q_{j}^{2}$ and, as before, integrate the resulting expression from $r_{b}-\epsilon$ to $r_{b}+\epsilon$. On the rhs, except for the last term all other terms give zero, and we get the 
second condition
\be 
\tan^{-1}\left[\frac{q_{j}}{p_{j}}\right]_{r_{b}-\epsilon}^{r_{b}+\epsilon}=-P,
\ee 
or defining $P=\tan^{-1}Q$, the above equation can be rewritten as 
\be \label{effect of delta potential 2}
\frac{q_{j>}}{p_{j>}}=\left(\frac{q_{j<}}{p_{j<}}-Q\right)\left(1+Q\frac{q_{j<}}{p_{j<}}\right)^{-1}. 
\ee 

Using the explicit solutions \eqref{q solution 2+1} and \eqref{p solution 2+1} and \eqref{effect of delta potential 2} in the $\epsilon\rightarrow0$ limit, the above condition can be rewritten as
\be \label{eigenvalue determining eq delta potential 2+1}
Q=-\frac{(b\pm\sqrt{b^{2}-x^{2}})}{x}\left[\frac{I_{j-\frac{1}{2}}(x)}{I_{j+\frac{1}{2}}(x)}+\frac{K_{j-\frac{1}{2}}(x)}{K_{j+\frac{1}{2}}(x)}\right]\left(1-\frac{(b\pm\sqrt{b^{2}-x^{2}})^{2}}{x^{2}}\frac{I_{j-\frac{1}{2}}(x)}{I_{j+\frac{1}{2}}(x)}\frac{K_{j-\frac{1}{2}}(x)}{K_{j+\frac{1}{2}}(x)}\right)^{-1}.
\ee
Note that in this equation the normalization factors $c_{1}$ and $c_{2}$ do not appear anywhere as only the ratio of the solutions are involved. As in the previous section, we have used the definition $m_{0}=b/r_{b}$ and $\epsilon_{j}r_{b}=x$.

Next we need to find the edge states from the above equation. As before, we find that the slope of the curves corresponding to the rhs of \eqref{eigenvalue determining eq delta potential 2+1} in the limit $x\rightarrow0$ is zero and, therefore, the same procedure can be used to find the number of edge states (and, as before, we only require $j>1/2$ for the argument to work). We start by noting that the two signs in $\pm\sqrt{b^{2}-x^{2}}$ correspond to positive and negative energy solutions. We therefore have four possibilities - (i) $E_{j}>0, j>0$ (ii) $E_{j}>0, j<0$ (iii) $E_{j}<0, j>0$ and (iv) $E_{j}<0, j<0$. Following the earlier method we find that these four possibilities will lead to two solutions each for $j_{\rm max}$ and $j_{\rm min}$. For instance, the first case above where we choose the positive square root and $j>0$, the total number of bound states is found to be
\be \label{number of edge states j>0 2+1 delta potential} 
j_{\rm max,1}=Qb+\frac{1}{2}.
\ee
Similarly, case two leads to 
\be \label{number of edge states j<0 2+1 delta potential} 
j_{\rm min,1}=-Qb+\frac{1}{2}. 
\ee 
Consistency with the requirement that $j_{\rm max,1}>0$ and $j_{\rm min,1}<0$ implies that $Q>0$. In a similar manner the other two cases, corresponding to choosing the minus sign for the square root, lead to $j_{\rm max,2}=-Qb-1/2$ and $j_{\rm min,2}=Qb-1/2$ and consistency implies that for these two cases $Q<0$.

Now the sign of $Q$ (equivalently of $P$) determines whether the particle sees a potential hill or a potential well. However, this interpretation also depends on whether the particle has positive energy or negative energy. The first two cases corresponding to positive energy and $Q>0$, therefore, imply that the particle sees a potential well. Similarly, the other two solutions for $Q<0$ correspond to a negative energy particle seeing a potential well. Thus we find that $j_{\rm max,1}=-j_{\rm min,2}$ and $j_{\rm max,2}=-j_{\rm min,1}$.

If we now equate the expression for one of the $j_{\rm max}$, say $j_{\rm max,1}$ in \eqref{number of edge states j>0 2+1 delta potential}, to the number found in \eqref{maximum number of edge states plus 2+1} in the presence of the APS boundary condition, we get the condition 
\be \label{choice for Q j>0}
Q=\frac{\sqrt{1+4A^{2}+8Ab}}{2(A+2b)},
\ee
which leads to the same number of edge states with $j>0$ for the problem with the delta potential as obtained in presence of the APS boundary condition. And as shown above, the negative of this $Q$ value leads to the same $j_{\rm min,2}$ as found with the APS boundary condition. It is interesting to see that the role of the sign of the potential is seen only for the edge states. The expression for $Q$ will be slightly modified for $j_{\rm max,2}$ (and $j_{\rm min,1}$) and is given by $Q=(2A+3b+b\sqrt{1+4A^{2}+8Ab})/2b(A+2b)$.

\section{Time dependent (moving) boundaries and boundary conditions}

The problem of quantum mechanics and quantum field theory on 
manifolds with time changing or moving boundaries is interesting 
in its own right but also has important physics implications. 
For instance, it might be of interest to study Casimir effect in 
the presence of moving boundaries and, in the context of general 
relativity and cosmology also such situations arise naturally 
in the form of evolving horizons (black hole or cosmological) 
which are often treated as boundaries. Moving mirror and associated 
Unruh effect are other examples of phenomena involving moving boundaries. 
In this section we, therefore, 
consider the question as to what happens to quantum 
mechanical problems when the boundary is changing with time? 
This change 
can be in variety of ways. There can be expansion, 
translation or there can be a change of shape of the boundary. 

The difficulty with quantum mechanical problems involving 
moving boundary is that the domain of the Hamiltonian 
operator in the space of ${\cal{L}}^2(M)$ (in order 
that the Hamiltonian be a self-adjoint operator) changes with time. 
As the boundary evolves, the wave function will go
outside the domain and hence the description of time evolution 
becomes subtle. To deal with this we have to deal with the 
whole Hilbert space and the expansion of functions in terms
eigenfunctions will change with time \cite{MarmoMovingWalls}.
 
\subsection{Expanding disc}

Below we consider the situation involving $\mathbb{R}^{2}-\mathcal{D}^{2}$ with the radius of the disc being a function of time and explore how to do quantum physics in such a situation.
We focus on the case where the Hamiltonian is Laplacian 
in the region $\mathbb{R}^2-\mathcal{D}^{2}$ and zero outside this domain 
(from the analysis of the previous sections it should be 
clear that situation involving the Dirac operator is 
not fundamentally different from that of the Laplacian).
  
The expanding radius is given by 
\be
R(t)=e^{f(t)}R_{0}, \qquad f(0)=0,
\ee
and we will be considering imposition of the Robin boundary condition with parameter $\kappa$.
The basic idea would be to translate the problem to one involving 
fixed boundary but a time dependent Hamiltonian. For this 
consider the following transformation of the wave function $\psi$:
\be
\chi(r,\theta)=U\psi(r,\theta),
\ee
where $U(r,\theta)=e^{-\alpha(r)}$.  
Then 
\be
\chi|_{\partial D}=e^{-\alpha}\psi(r,\theta)|_{\partial D}.
\ee 
On the other hand, the normal derivative of $\chi$ is (remember that we are imposing Robin boundary condition)
\be
\partial_n \chi|_{\partial D}=-\partial_n \alpha|_{\partial D}
\left(e^{-\alpha} \psi(r,\theta)\right)|_{\partial D}-
e^{-\alpha(r)}\kappa\psi(r,\theta)|_{\partial D}=-(\kappa +\partial_
{n}\alpha)\chi|_{\partial D},
\ee
where we have made use of $\psi'(r)|_{\partial D}=-\kappa\psi(r)|_{\partial D}$.
Thus we see that the transformation effects a change in the boundary condition
(and thereby the domain in the Hilbert space) by changing 
$\kappa\rightarrow\kappa'=\kappa+\partial_n\alpha|_{\partial D}$ (note that the prime on $\kappa$ does not denote derivative with respect to $r$).

Now the above transformation implies that the Hamiltonian changes to $H'=UHU^{-1}$.
Using this we can maintain the domain of the Hilbert space same 
while changing the Hamiltonian and hence the spectrum. If the 
$\alpha$ depends on time the Hamiltonian will change to 
\be
H'=UHU^{-1}+i\frac{\partial U}{\partial t}U^{-1}.
\ee
It will be a time dependent Hamiltonian.  

We want that $\kappa'$, the new parameter for the Robin boundary condition when the problem is translated to one with a fixed boundary $R(0)$, be such that the physics for the problem $(\kappa',R(0))$ is the same as that for $(\kappa,R(t))$. In particular, this means that the number of edge states for the two problems should be the same. As was shown in \cite{TRG-Novel}, the number of edge states for the Laplacian on $\mathbb{R}^{2}-\mathcal{D}^{2}$ is $-\kappa r_{b}$ ($r_{b}$ being the radius of the boundary) and this implies that
\be
\kappa'R(0)=\kappa R(t).
\ee
Hence $\kappa'=\frac{R(t)}{R(0)}\kappa=e^{f}\kappa$. 
But we also know:
\be
\partial_n\alpha=\kappa'-\kappa.
\ee
Hence we get
\be
\alpha=(\kappa'-\kappa)r=(e^{f}-1)\kappa r.
\ee
Having obtained $\alpha(r,t)$ we can write $U$ explicitly
and hence the time dependent $H'$. The transition amplitudes 
are the important quantities that can be found using the 
Hamiltonian which is time dependent.
  
\subsection{Delta function potential}
It is well known that in one dimension the boundary conditions 
can be obtained from a delta function potential and, in the 
previous section, we already demonstrated that the same holds 
even in higher dimensions. In principle the strength of the 
potential can vary along the boundary circle 
so that boundary conditions 
have angular dependence. We will ignore this 
possibility and take the strength 
to be uniform along the boundary circle.

Consider the following Hamiltonian in $\mathbb{R}^2$ 
\be 
H=-\Delta+g\delta(r-R(t)).
\ee
Here the delta function along a curve ${\cal{C}}$ is defined as:
\be
\int \delta(f({\cal{C}})) \psi(x) d^2x=\int_{\cal{C}} \psi(x) dx.
\ee
For example when $\psi(x)=1$ and curve $\mathcal{C}$ is a circle of radius $R$ we get $2\pi R$.
Now $R(t)=e^{f}R(0)$ and if we apply dilation operator $D$ such that 
$r \rightarrow kr$ then we get for the potential 
$\delta(kr-kR(t))$. If we fix $k$ to be $e^{-f}$, the potential 
becomes $\delta(e^{-f}r-R(0))=e^{f}\delta(r-R(0))$. The 
time dependence is shifted to the strength of potential. This is
analogous to changing the Hamiltonian to a time dependent 
one by keeping the domain of the Hilbert space same for all times.

\subsection{Solvable time dependent boundary problems}
Interestingly, the case of one dimensional delta function potential 
whose strength varies with time can be exactly solved in several
cases. Several of these time dependent potentials correspond to 
problems with moving boundaries, for example, where the boundary moves with uniform
velocity or has uniform acceleration. In the first case the Galilean 
invariance provides the solution. In the second case the Hamiltonian 
can be transformed to a constant gravitational field whose solution 
is given by Airy functions \cite{MarmoMovingWalls, BerryKlein}.  

There are several examples in higher dimensions where the time dependent 
problem can be exactly solved which corresponds to moving boundaries.
One such example is given by Berry and Klein \cite{BerryKlein}:
\be
H(r,p,l(t))=\frac{p^2}{2m}+\frac{1}{l^2}V(r/l).
\ee
Here $l$ depends on time. This Hamiltonian is not conserved.
If the time dependence is of the form $l(t)=\sqrt{at^{2}+2bt+c}$
then one can write in a comoving frame 
\be
H(\rho,\pi,k)=\frac{\pi^2}{2m}+V(\rho)+\frac{1}{2}k\rho^{2},
\ee
where $\rho=r/l$ and $k=m(ac-b^{2})$ which is conserved 
in $\rho,\tau\equiv\int^{t}\frac{dt}{l^{2}(t)}$. 
The expanding disc in $\mathbb{R}^2$ and ball in $\mathbb{R}^3$ will come under 
this class of Hamiltonians. A detailed analysis of the quantum mechanics
with implications for field theory as well as extension to Dirac
Hamiltonian will be presented elsewhere.

\section{Applications} Manifolds with boundary is considered 
in several applications. We will list few of them. Analysis 
in several new situations will be presented elsewhere.  
\begin{itemize}
\item
Quantum Hall effect: This is a classic example of electron gas 
in a two dimensional manifold with boundary. Naturally the edge 
excitations are chiral and has been analysed to a large degree. 
Topological insulators arising in 2d and 3d also can be understood 
in this context.  See for example \cite{BalChiralBags}. But moving boundaries 
in this context have not been considered both theoretically and 
experimentally.

\item
Casimir effect: Casimir effect of force between plates due 
to vacuum fluctuations can be analysed using self-adjoint extensions
of the Laplacian or Dirac operators with suitable boundary conditions. Effects of modelling the plates in terms of singular potentials has been investigated in several recent papers \cite{BartonPlasmaShells, GuilarteCastanedaDoubleDelta, ParasharDeltaPlate, CastanedaMosqueraCasimirDelta, CastanedaGuilartedeltadelta'}. Asorey et al \cite{AsoreyCastanedaGenBoundCond} have analyzed the effect of boundary conditions on the Casimir force and have showed that depending on the nature of the boundary condition this force can be attractive, repulsive or it can also vanish. The focus is on positive Laplacian operator and it would be interesting to see the effect of finite temperature and self-interactions.  
Detailed changes in the nature of the boundary like half-line 
or boundary with holes have been considered in the literature
\cite{NairCasimir}. The nature of this effect with moving boundaries 
will be new tool for testing  the developments posed here.

\item
Isolated and cosmological horizons: Blackholes 
present novel and unstable behaviour classically due to the 
appearance of null surfaces such as horizons. 
They require quantum theory to understand several questions. 
Membrane paradigm \cite{DamourEddy, ThorneMembraneParadigm} is one framework in which the horizon
acts as boundary and requires boundary conditions. One can 
analyse the situation using self-adjoint extensions 
for suitable boundary conditions. Similar situation arises 
for the cosmological horizon of de Sitter space of evolving universe.
The nature of evolving boundaries in this situation can be used to 
relate the stationary coordinates and ingoing observers.

\item Edge excitations localised in the boundary have dramatic effects
in gas of bosons as pointed out by one of us \cite{trgnair}. Evolving
boundaries in this case and effects on statistical mechanics of gas 
of bosons/fermions require further analysis. A nice example of edge states
actually appearing and requiring clear understanding
of boundary conditions is provided by plasmons \cite{HanEdgeStatePlasmonic, ChristensenClassQuantPlasmonics, LingTopoEdgePlasmon}.   
\end{itemize}

\section{Discussion}

Problems requiring solution of quantum mechanical equations on manifolds with boundaries feature very often in physics. The most important issue in these situations is that of the self-adjointness of the corresponding Hamiltonian operator on the given domain. Although the two most commonly employed boundary conditions in the physics literature are the Dirichlet and the Neumann boundary conditions, it is known that self-adjointness of the Hamiltonian permits more general boundary conditions. For the case of the Laplacian these are the Robin boundary conditions while for the Dirac operator these are the APS boundary conditions.

In this paper we showed that the APS boundary condition applicable for the Dirac operator can be thought of as generalized Robin boundary condition. The Robin boundary condition is usually given in the form $\psi'(r)|_{\partial D}=-\kappa\psi(r)|_{\partial D}$ where $\kappa$ is a dimensionful parameter and is usually taken to be a constant. If we now look back at the equation \eqref{generalized boundary condition 2+1} which implements the APS boundary condition in $2+1$ dimensions (and the corresponding equation \eqref{boundary condition generalized boundary operator 3+1} for $3+1$ dimensions) then it is obvious that these are essentially of the same form as the Robin boundary condition.

However, there is an essential difference. While in the usual application of the Robin boundary condition the parameter $\kappa$ is taken to be a constant, we saw that, more generally, the parameter can depend on the spin (angular momentum) state $j$ as well on the corresponding energy eigenvalue in the given state (in addition to the mass parameters $m_{0}$ and $\mu$). We gave an exact counting for the number of edge states for the Dirac operator in presence of the APS boundary condition in $2+1$ dimensions as well as in $3+1$ dimensions for the case when the boundary has the isometry of a sphere.

It is known that the issue of self-adjointness of operators becomes much more intricate in the presence of boundaries. One also expects that in actual physical situations the boundaries and boundary conditions would be implemented by suitable interactions. Based on this idea we showed that the effect of boundary conditions can be mimicked by extending the spatial manifold beyond the boundary (so that there is no boundary any more) and by putting a delta function potential of suitable strength at the location of the boundary. We worked out this correspondence explicitly for both, the Laplacian and the Dirac operator. For the case of the Laplacian we showed how the strength of the delta potential is related to the parameter $\kappa$ appearing in the Robin boundary condition, so that both the problems lead to the same number of edge states. 

For the case of the Dirac operator the correspondence is essentially there except for one caveat that while in the case of the APS boundary condition the number of edge states with $j>0$ is the same as the number of states with $j<0$ (for a given choice of the parameters), for the case where the boundary condition is replaced by a delta potential, the number of edge states (for a given value of the strength of the delta potential) for $j>0$ is one more than that for $j<0$. Here it is also useful to note that like the parameter $\kappa$ of the Robin boundary condition, the strength of the delta potential is a dimensionful quantity in the case of the Laplacian (a second order operator) whereas for the Dirac operator, the strength of the delta potential is dimensionless since the Dirac operator is a first order operator.

Finally we showed how the complicated problem of moving boundaries can be translated to a different form. In the presence of moving boundaries, the domain of the Hamiltonian changes with time and therefore the question of self-adjointness of operators becomes much more complicated \cite{MarmoMovingWalls}. We showed that it is possible to make a transformation such that the boundary becomes fixed and the effect of the moving boundary is contained in a modified Hamiltonian and boundary condition.

As previously alluded to, their are numerous situations where one is required to do quantum mechanical calculations on manifolds with boundaries (including moving boundaries). These include quantum Hall systems and topological insulators on the condensed matter side and cosmological and black hole horizon on the gravity side. For instance, use of self-adjoint extensions for the Dirac Hamiltonian on graphene with defects was considered  in \cite{StoneFullereneSelfAdjoint}. Additionally, interesting
question regarding heat kernel expansion for scalar field theory on finite interval has been analyzed in  \cite{KirstenQFTfiniteLine} following the earlier work in \cite{DowkerHeatKernelCone}.  As in \cite{BordagVassilevicHeatKernel, BordagVassilevicNonSmoothQFT},  only positive
Laplacian is considered to obtain heat kernel coefficients. It will be interesting to extend this analysis to finite temperature condition
or with self interactions. In a future work we hope to investigate the implications of the analysis of this paper on some of these physical systems.

\section*{Acknowledgements}
TRG acknowledges M. Berry, A.P. Balachandran, Giuseppe Marmo and Manuel Asorey for discussions. RT acknowledges the hospitality at Chennai Mathematical Institute where part of this work was done. The authors thank the referee for his/her comments as well as for suggesting several useful references. The research of RT is supported under the DST/1100 project of the Department of Science and Technology, India.

\end{document}